\documentclass[12pt]{article} 
\usepackage[utf8]{inputenc}
\usepackage{multirow}
\usepackage{amsfonts}
\usepackage{amsmath}
\usepackage{amssymb}
\usepackage{amsthm}
\usepackage{caption}
\usepackage[dvipsnames]{xcolor}
    \definecolor{darkgreen}{rgb}{0,0.5,0}
    \definecolor{darkblue}{rgb}{0,0,0.6}
    \definecolor{purple}{rgb}{0.4,.2,0.7}
\usepackage[margin = 2.5cm]{geometry}
\usepackage{graphicx}
\usepackage[unicode=true,
 bookmarks=true,bookmarksnumbered=false,bookmarksopen=false,
 breaklinks=false,pdfborder={0 0 1},backref=false,colorlinks=true]
 {hyperref}
\usepackage{subcaption}

\newcommand{\be}{\begin{equation}}
\newcommand{\ee}{\end{equation}}
\newcommand{\bea}{\begin{eqnarray}}
\newcommand{\eea}{\end{eqnarray}}
\def\la{\label}
\def\nref#1{(\ref{#1})}

  \hypersetup{pdfauthor={Juan Maldacena, Grant N. Remmen},
 citecolor=blue,linkcolor=black,urlcolor=black}

\usepackage{cite}
\usepackage{breakurl}
\usepackage[hang,flushmargin]{footmisc}

\begin{document}

\hfill 

\interfootnotelinepenalty=10000
\baselineskip=18pt
\hfill

\vspace{1.36cm}
\thispagestyle{empty}
\begin{center}

{\LARGE \bf
Accumulation-Point Amplitudes \\ in String Theory
}
\\

 \vspace{1.36cm}
{
{\large Juan Maldacena${}^{a}$ and Grant N. Remmen${}^{b,c}$}
} \\[7mm]
{
\it ${}^a$Institute for Advanced Study,  Princeton, NJ 08540 \\[1.5mm]
${}^b$Kavli Institute for Theoretical Physics, \\[-1mm]
University of California, Santa Barbara, CA 93106\\[1.5mm]
${}^c$Department of Physics, \\[-1mm]
University of California, Santa Barbara, CA 93106 \\[1.5 mm]}
\let\thefootnote\relax\footnote{e-mail:
 \url{malda@ias.edu}, \url{remmen@kitp.ucsb.edu}}

 \end{center}

\bigskip
\centerline{\large\bf Abstract}
\begin{quote} \small
We point out some common qualitative features of the Coon amplitude---a family of deformations of the Veneziano amplitude with logarithmic Regge trajectories---and the open string scattering amplitude for strings ending on a D-brane in AdS. Both reduce to the Veneziano amplitude at relatively low energies. Both systems have an accumulation point in their spectrum,  with an infinite number of states below a certain energy. The approach to this point is very similar. Both have the same high-energy  fixed-angle behavior. Nevertheless, we find some differences in the spectrum of states with highest angular momentum. 
These similarities suggest that there may exist a string background that realizes the Coon amplitude.

\end{quote}
	
\setcounter{footnote}{0}

\newpage

\section{Introduction}

In 1969, Coon~\cite{Coon:1969yw} constructed a one-parameter deformation of the Veneziano amplitude that appears to correspond to a unitary S-matrix~\cite{Figueroa:2022onw}. 
While the Veneziano amplitude famously describes a spectrum with linearly increasing mass-squares, the Coon amplitude exhibits an accumulation point, 
with an infinite number of states below a certain energy,  leading to questions about its physical relevance \cite{Caron-Huot:2016icg}. 
Furthermore,  somewhat similar infinite towers of states have been found in recent amplitude studies~\cite{Caron-Huot:2016icg,Bern:2021ppb,Huang:2022mdb,Arkani-Hamed:2020blm}. 

In this paper, we point out that the scattering amplitude for open strings on D-branes in ${\rm AdS}$ has several features in common with the Coon amplitude, including the accumulation point in the spectrum and the high-energy behavior. We do find some differences in the behavior of the large angular momentum states. We hope that some of our remarks below will be useful for finding the correct string background leading to the Coon amplitude, if one exists!

This paper is structured as follows. In Sec.~\ref{sec:Coon}, we review the Coon amplitude and the important features that for our analysis.
In Sec.~\ref{sec:string}, describe the D-brane construction in ${\rm AdS}$ that leads to a similar amplitude.  
Finally, in Sec.~\ref{sec:comparison}, we compare aspects of the spectra and amplitudes in the Coon and string cases.

\section{The Coon amplitude } \label{sec:Coon}

The Coon amplitude~\cite{Coon:1969yw} is a modification of the standard Veneziano amplitude for the scattering of open strings; see Ref.~\cite{Figueroa:2022onw} and references therein. It reads\footnote{Here we write in the Coon amplitude with the exponential prefactor needed for locality~\cite{Figueroa:2022onw} and using the standard sign convention for scattering amplitudes. Unitarity requires $q \in [0,1]$ ($\gamma \geq 0$).}
\be 
{\cal A}_{\rm C}(s,t) = g^2 (1-q) \exp\left({\frac{\log \sigma \log \tau}{\log q}} \right) \prod_{n=0}^\infty \frac{\left(1-\frac{q^n}{\sigma\tau}\right)(1-q^{n+1})}{\left(1-\frac{q^n}{\sigma}\right)\left(1-\frac{q^n}{\tau}\right)}\label{eq:AC},~~~~~q = e^{-\gamma},
\ee 
where $\sigma = 1 + (\alpha_0 + \alpha' s)(q-1)$ and $\tau = 1 + (\alpha_0 + \alpha' t)(q-1)$, and $g$ is a  coupling constant.
The amplitude contains a dimensionless parameter $q$ that for later convenience we have written as $e^{-\gamma}$.   In the $q\rightarrow 1$ ($\gamma\rightarrow 0$) limit we recover the Veneziano amplitude,
\be 
{\cal A}_{\rm V}(s,t) = g^2 \frac{\Gamma(-\alpha_0-\alpha' s)\Gamma(-\alpha_0 - \alpha' t)}{\Gamma(-2\alpha_0 + \alpha' u)}.
\ee 
We will be mostly interested in the regime where $\gamma$ in Eq.~\nref{eq:AC} is small, so that $q \sim 1$. We will also later assume that $\alpha_0 =1$, as in the usual Veneziano case. 
Looking at the poles of Eq.~\eqref{eq:AC} at $s,t=m_n^2$, we find that the spectrum is given by 
\be 
\alpha' m_n^2 = -\alpha_0 + \frac{1-q^n}{1-q}.
\ee 
We see that the Coon amplitude has an accumulation point 
that is approached exponentially, 
\be \la{AccCoon}
\alpha'(m_n^2 - m_\infty^2) = -{ q^n \over 1-q} , ~~~~~~~~~~   \alpha' m^2_\infty = { 1 \over 1\,{-}\,q }-\alpha_0 \sim { 1 \over \gamma } .
\ee 
For each pole labeled by $n$, there is a tower of states with spin ranging from $0$ to $n$.

When $\gamma \sim 0$, at relatively low energies the Coon amplitude agrees with the low-energy Veneziano amplitude. For larger energies we transition into the high-energy behavior of the Veneziano amplitude, and ultimately when $\alpha' s \gamma \sim O(1)$ we transition further into the true high-energy behavior of the Coon amplitude,
\be \la{HighC}
{\log {\cal A}_{\rm C}} \sim  -\frac{\log s \log t}{\gamma} + \cdots ,~~~~~~{\rm for} ~~~~ t, s \gg {  1 \over \alpha' \gamma}, 
\ee 
which results from the exponential prefactor in Eq.~\nref{eq:AC}.
In  appendix~\ref{App:High}, we give the high-energy behavior for all values of $s \gamma$.

\section{Open string amplitudes for branes in AdS}\label{sec:string}

\begin{figure}[t]
    \begin{center}
    \includegraphics[height=8cm]{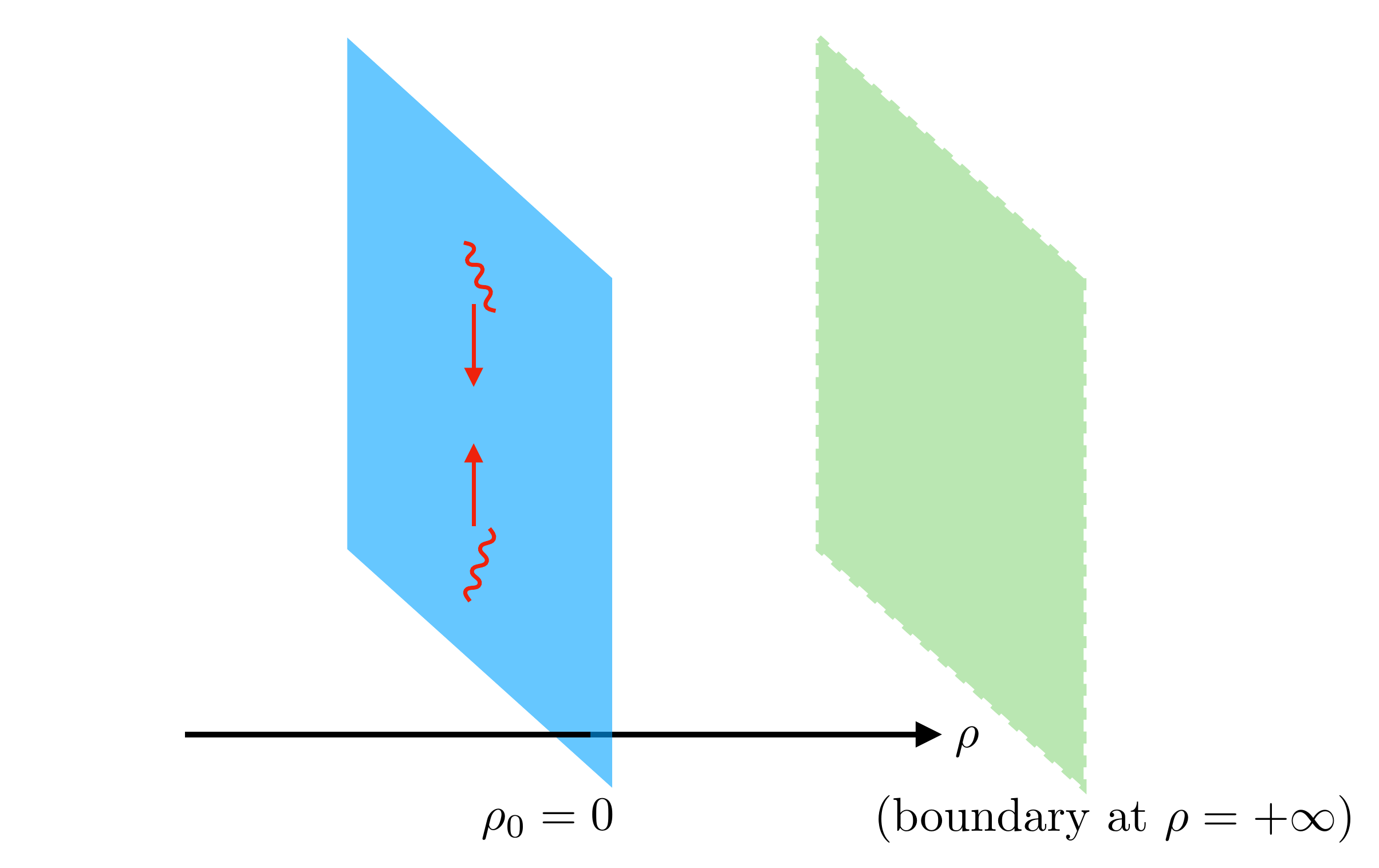}
    \end{center}
    \caption{Scattering of open strings in AdS, attached to a D-brane (blue) at coordinate $\rho = 0$, for the metric given in Eq.~\nref{Met}.}
    \label{AdSScattering}
\end{figure}

We now consider an open string scattering problem that shares features of the Coon amplitude.
To this end, let us consider an ${\rm AdS}_{d+1}$ metric of the form 
\be \la{Met}
{\rm d}s^2 =  {\rm d}\rho^2 + e^{2 \rho/R} {\rm d}x_d^2,
\ee 
where $d\geq 3$, $R$ is the ${\rm AdS}$ radius, and ${\rm d}x_d^2$ is the flat $d$-dimensional metric. 
We place a D-brane at $\rho = \rho_0$, extended along all $d$ spacetime directions labeled by $x^\mu$. By scaling symmetry, any choice of $\rho_0$ has the same physics, so we choose $\rho_0=0$ so that the proper time at this position agrees with the coordinate $x^0$. 
We will consider the scattering of open strings attached to the D-brane, as depicted in Fig.~\ref{AdSScattering}.\footnote{ For the D-brane to be an actual solution we need more ingredients, such as  Ramond-Ramond fluxes in a type II superstring theory. We will ignore such issues here because our analysis will be insensitive to these details.}

We analyze this problem in the large-radius limit, where   
\be \la{smallalp}
\tilde \gamma \equiv { \alpha' \over R^2 }  \ll 1.
\ee 
In this limit, for relatively low energies satisfying
$\alpha' s \tilde \gamma \ll 1 $, we can approximate the scattering by the flat-space Veneziano amplitude. Note that due to Eq.~\nref{smallalp}, this regime includes the region $\alpha' s \gg 1$.  
 As we review in appendix~\ref{DevString}, at high energies the amplitude  has the schematic form 
 \be \la{HSgen}
 \log {\cal A}_{\rm String} = - { 1 \over \tilde \gamma } F(\alpha' s \tilde \gamma , \alpha' t \tilde \gamma ) ,~~~~~{\rm for } ~~ t, s \to \infty, ~~\tilde \gamma \to 0,~~~~{\rm with} ~~t/s,~s \tilde \gamma  ~~{\rm fixed.} 
 \ee 
 This behavior arises from a classical string solution in ${\rm AdS}$ \cite{Alday:2007hr}, as in the flat-space analysis of Ref.~\cite{Gross:1987ar}. 
  When $\alpha' s \tilde \gamma $ is small, we obtain the high-energy behavior of the Veneziano amplitude, while for large values of $\alpha' s \tilde \gamma $ we find~\cite{Alday:2007hr}
  \be \la{HighS}
  \log { \cal A}_{\rm String} \sim -\frac{\log s \log t}{4\pi \tilde \gamma} + \cdots ,~~~~~{\rm for }~~  s , t \gg { 1 \over \alpha' \tilde \gamma } .
 \ee

 \begin{figure}[t]
    \begin{center}
    \includegraphics[height=8cm]{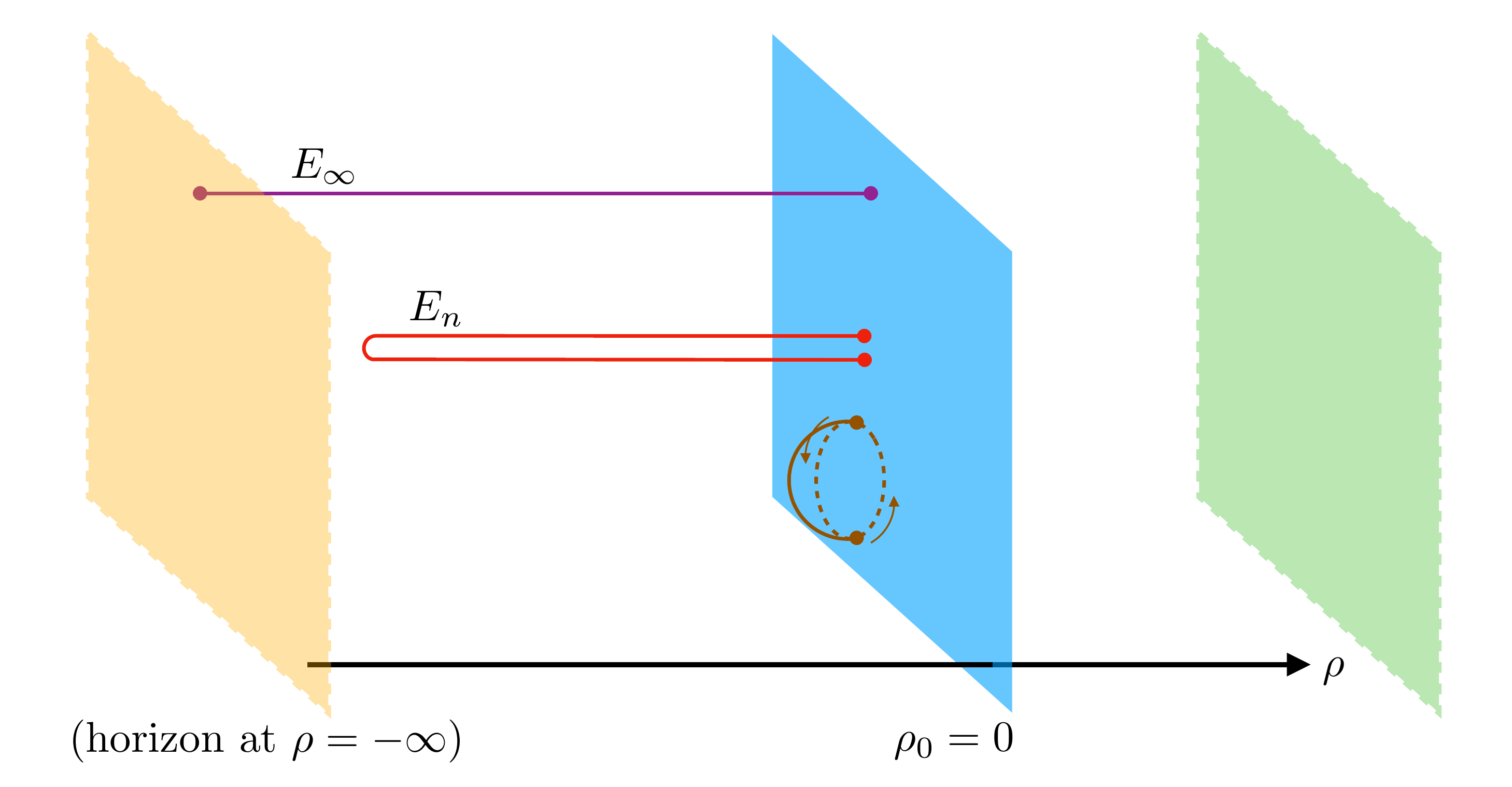}
    \end{center}
    \caption{Various string configurations discussed in the text.
In purple, we  depict a string stretching between the brane (blue) and the horizon (yellow). 
A pair of such strings has the energy of the continuum threshold. In red,  we show a state of a lower energy consisting of a folded string that stretches up to a finite distance from the brane, with the endpoint momentarily at rest. 
This endpoint will start moving towards the brane.  
Finally, the spinning string is depicted in brown.}
    \label{VariousStrings}
\end{figure}
 
 The spectrum contains an accumulation point because we can consider a string that stretches from $\rho=0$ to $\rho = -\infty$; see Fig.~\ref{VariousStrings}. Such a string has  a finite energy that we will write as $E_\infty/2$, where
 \be \la{Ega}
 \frac{E_\infty}{2} = { 1 \over 2 \pi \alpha' } \int_{-\infty}^0 {\rm d}\rho\, e^{\rho/R} = { R \over 2 \pi \alpha' } = { 1 \over 2\pi \sqrt{\alpha'} } { 1 \over \sqrt{\tilde \gamma } } .
 \ee 
When the center-of-mass energy $s$ is larger than $E_\infty^2$, we can produce a continuum of states consisting of a pair of these stretched strings moving away from each other.

 It is interesting to consider the string spectrum for energies lower than Eq.~\nref{Ega}. A particular set of string states is obtained by considering a folded string stretching from $\rho=0$ to a negative value $\rho_0$. For large $|\rho_0|$ it will have an energy slightly less than twice Eq.~\nref{Ega}. If we have a folded string of this kind at some instant, it would start approaching the brane at later times. We can semiclassically quantize this motion, and one finds a set of states given by Ref.~\cite{Klebanov:2006jj},
 \be 
\la{AccString}
  E_n = E_\infty\left(1 -  c~e^{ - { \pi^2 \alpha' \over R^2 }  n } \right),~~~~~{\rm for}~~n \gg 1,
    \ee 
where $c$ is an $O(1)$ constant that we will not compute. 
As in the case of the Coon spectrum, here we are approaching the threshold energy in an exponential manner. This exponential $n$-dependence is largely determined by the conformal symmetry of ${\rm AdS}$ \cite{Klebanov:2006jj}. 
 
 It is also interesting to compute the form of the energy for the lowest-energy rotating string attached to the brane; see Fig~\ref{VariousStrings}. We discuss these solutions in more detail in appendix~\ref{SpinningStrings}, in the classical limit.
  For relatively small angular momenta, this spectrum is given by the flat-space formula,  
\be 
\alpha' E^2 = J.\label{eq:lowJ}
\ee 
For large angular momenta, the spectrum can be computed by treating the string as a pair of massive particles interacting via a Coulomb-like potential. The mass of each of these particles is $E_\infty/2$ in Eq.~\nref{Ega}. The potential is obtained from the same computation that gives the quark-antiquark potential in Ref.~\cite{Maldacena:1998im}.  This gives
 \be 
E - E_\infty  = -\frac{2\pi^3}{\Gamma(1/4)^8 R} \left(\frac{R^2}{\alpha'}\right)^3\frac{1}{J^2}.\label{eq:highJ}
 \ee 

\section{Comparison between the two amplitudes}\label{sec:comparison}

The Coon amplitude and the string amplitude for a string in ${\rm AdS}$ have some qualitative features in common. Both contain a small parameter, which we have denoted by $\gamma$ and $\tilde \gamma$ in Eqs.~\nref{eq:AC} and \nref{smallalp}, respectively. For relatively small energies, we recover the low-energy behavior of the Veneziano amplitude. For larger energies we find a deviation from the Veneziano amplitude. 
In both cases we see a spectrum with an (exponentially-approached) accumulation point, as shown in Eqs.~\nref{AccCoon} and \nref{AccString}.
In both cases, the very high-energy behavior (at fixed angles) is similar, as we find in Eqs.~\nref{HighC} and \nref{HighS}. 

One could wonder whether there is a choice for the relation between $\gamma$ and $\tilde \gamma$ that would make the Coon and AdS string amplitudes match precisely. If we take $\gamma = \pi^2 \tilde \gamma$, then the threshold masses---$m_\infty$ and $E_\infty$ in Eqs.~\nref{AccCoon} and \nref{AccString}, respectively---as well as the approach to the accumulation point---given by the factor multiplying $n$ in the exponential---do indeed match. However, then the high-energy behaviors of $\log {\cal A}$ in Eqs.~\nref{HighC} and \nref{HighS} differ by a multiplicative factor. 

On the other hand, the spectrum of rotating strings is different at large $J$. In the Coon amplitude, the minimum energy of a state with a given angular momentum corresponds to setting $n = J$ in Eq.~\nref{AccCoon}, and it approaches the accumulation point exponentially. However, in the situation of the brane in ${\rm AdS}$, the spectrum for large $J$ is given by Eq.~\nref{eq:highJ} which approaches the accumulation point as $1/J^2$. 

The discussion in this paper suggests that the Coon amplitude might correspond to ordinary strings in a suitable background. It is not precisely the ${\rm AdS}$ background we discussed here,  because we did not find perfect agreement.\footnote{ We could also consider the  $\tilde \gamma \gg 1$ limit, which  in the case of ${\rm AdS}_5 \times S^5$ is dual to a weakly coupled gauge theory.  We did not find a regime in the Coon amplitude that would give a qualitative agreement in this case.} We hope that this discussion will be useful for finding the appropriate background, if any exists.

\vspace{5mm}
 
\begin{center} 
{\bf Acknowledgments}
\end{center}
\noindent 
We thank C. Cheung and A. Zhiboedov for useful discussions and comments. 
J.M. is supported in part by U.S. Department of Energy grant DE-SC0009988 and by the Simons Foundation grant 385600.
G.N.R. is supported at the Kavli Institute for Theoretical Physics by the Simons Foundation (Grant No.~216179) and the National Science Foundation (Grant No.~NSF PHY-1748958) and at the University of California, Santa Barbara by the Fundamental Physics Fellowship.

\appendix
\section{The high-energy limit of the Coon amplitude } 
\la{App:High}

In this appendix,  we consider the Coon amplitude in the fixed-angle, high-energy, small-$\gamma$ limit,
\be \la{HiLim} 
 \gamma \to 0 ,~~~~~s, t \to \infty,~~~~~{\rm  with} ~~ t/s ~~{\rm and} ~~ { s   \gamma } ~~ {\rm fixed.}
 \ee 
  In addition, we take $s, t < 0$ for simplicity. 

The limit can be computed simply by replacing the products in Eq.~\nref{eq:AC} by sums using 
\be 
\log\left[\prod_{n=1}^\infty \left( 1 - { e^{ - n \gamma} \over a } \right)\right] \simeq  \int_0^\infty {\rm d}n \log \left(1 - { e^{ -\gamma n } \over a } \right) = -{ 1 \over \gamma }   {\rm Li}_2\left({ 1 \over  a } \right),
\ee 
which holds in the limit of small $\gamma$ for $a\geq 1$.
Using this formula for each of the four terms in the product in Eq.~\nref{eq:AC} and then taking the limit \nref{HiLim}, we find 
\be 
\begin{aligned}
\log {\cal A}_{\rm C} &\sim - { 1 \over \gamma} \bigg\{ 
\text{Li}_2\left[\frac{1}{(1-  s\gamma
    ) (1-  s\gamma  )}\right]-  \text{Li}_2\left(\frac{1}{1- s \gamma}\right)-\text{Li}_2\left(\frac{1}{1-  t \gamma}\right) \\ 
   & \qquad\qquad +\log
   (1- s \gamma ) \log (1- t\gamma )+\frac{\pi ^2}{6} \bigg\},
\end{aligned}
   \ee 
where we set $\alpha'=1$. When $s \gamma $ is small, we reproduce the standard high-energy limit of the Veneziano amplitude,
\be 
\log {\cal A}_V \sim  - s \log(-s) - t \log(-t) + (s + t) \log (-s-t) ,
\ee 
while in the limit of large $s\gamma$, we find the scaling in Eq.~\nref{HighC}.

\section{The high-energy limit of the string amplitude in AdS}\label{DevString}

In this appendix, we briefly review the solutions describing the high-energy scattering amplitudes for open strings anchored on a D-brane in ${\rm AdS}$, as discussed in Ref.~\cite{Alday:2007hr}. High-energy string amplitudes can be computed using a classical solution~\cite{Gross:1987ar}. We write the metric as ${\rm d}s^2 = ({\rm d}x^2 + {\rm d}z^2)/z^2$ and put the brane at $z=1$.  The solution involves four open strings with large momenta $k^\mu$. In order to visualize the solution, it is convenient to introduce T-dual worldsheet coordinates defined by ${\rm d} y = *  { {\rm d} x  \over z^2 }$ and also define $r = 1/z$. After these definitions,  we have a problem where we are searching for a minimal surface in the space ${\rm d}s^2 = ({\rm d}y^2 + {\rm d}r^2)/r^2$ that ends at $r=1$ on a polygon with null sides, with each side being equal to the spacetime momentum $k^\mu$. The action is proportional to the string tension, which gives the overall factor $R^2/\alpha' = 1/\tilde \gamma$ in Eq.~\nref{HSgen}, with $F$ in that equation being a function of the shape of the polygon; this shape depends on $s$ and $t$, which are proportional to the square of the distances between opposite vertices of the quadrilateral.

\section{Spinning string construction} \label{SpinningStrings}

In this appendix, we construct the relation between energy and angular momentum for a classical spinning string anchored to a D-brane in ${\rm AdS}$.
As discussed in text, we consider an ${\rm AdS}_d$ spacetime, with a D-brane at fixed $z$, where the ${\rm AdS}$ metric is ${\rm d}s^2 = R^2({\rm d}z^2 + {\rm d}x_d^2)/z^2$. (In this appendix, we use the coordinate $z \propto e^{-\rho/R}$ and will write $x^0$ as $t$.) 
The worldvolume of the brane occupies the Poincar\'e slice of ${\rm AdS}$, which we choose to locate at $z=1$ without loss of generality.
Starting with the Polyakov action describing the string worldsheet theory,
\be
S= - \frac{1}{4\pi \alpha'} \int {\rm d}^2 \sigma \sqrt{-h} h^{ab} g_{\mu\nu}[X] \partial_a X^\mu \partial_b X^\nu,
\ee
let us choose conformal gauge in which the worldsheet metric $h_{ab}$ is flat. 
The Virasoro constraints enforcing the vanishing of the stress-energy associated with $h$ become the requirements that $g_{\mu\nu} \dot X^\mu X'^\nu = 0$ and $g_{\mu\nu}(\dot X^\mu \dot X^\nu + X'^\mu X'^\nu) = 0$, writing $\dot{} = \partial_\tau$ and ${}^\prime = \partial_\sigma$ for worldsheet coordinates $(\tau,\sigma)$.
The equation of motion for $X^\mu$ is given by $\ddot X^\mu  + \Gamma^\mu_{\nu\rho}[g]\dot X^\nu  \dot X^\rho  -X''^\mu - \Gamma^\mu_{\nu\rho}[g]X'^\nu X'^\rho =0$.

\begin{figure}[t]
\begin{center}
\includegraphics[width=11cm]{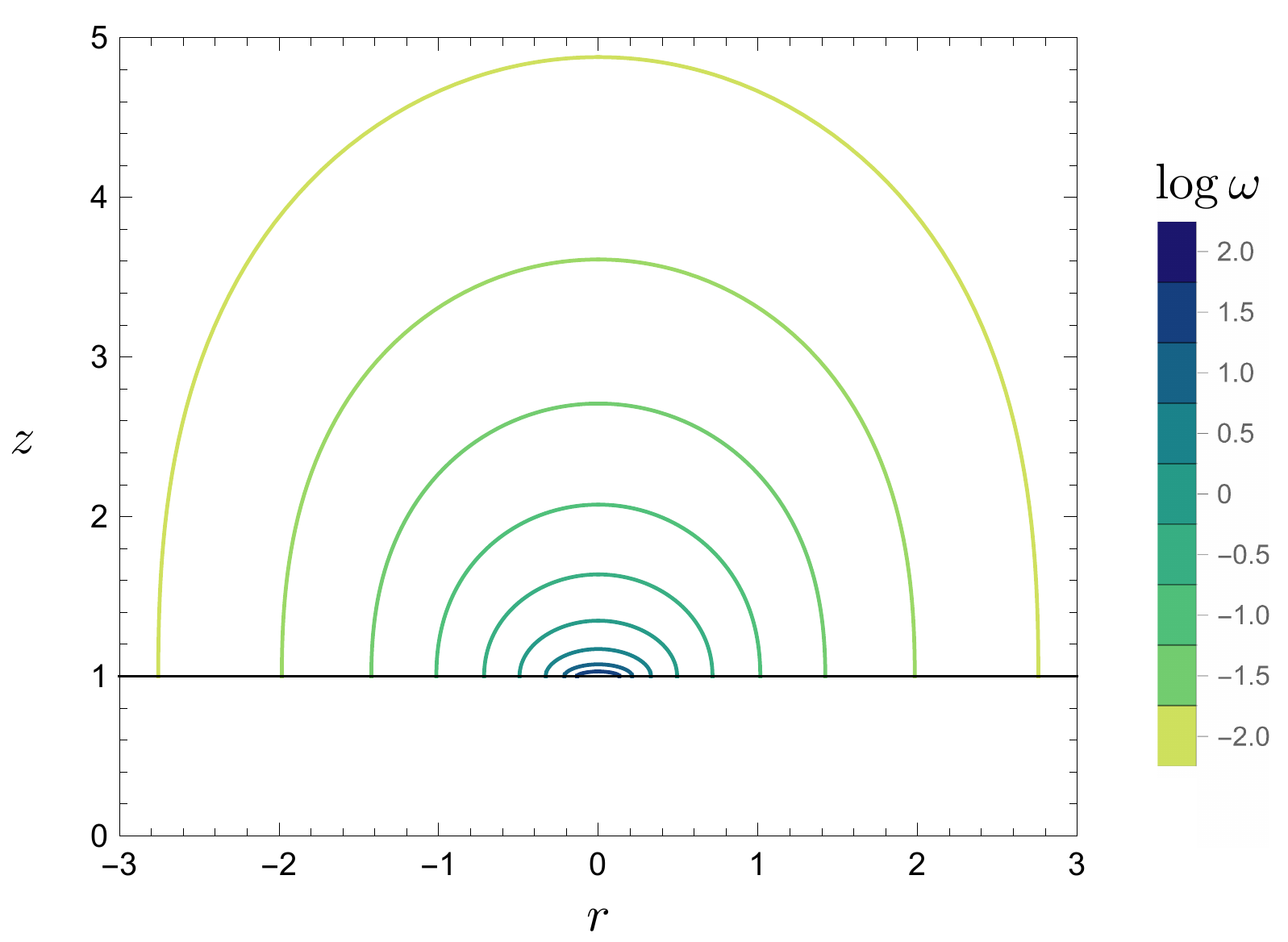}
\end{center}
\vspace{-0.5cm}
\caption{Profile of rotating strings attached to a D-brane in ${\rm AdS}$ for small-$\omega$/large-$J$ (yellow) to large-$\omega$/small-$J$ (blue), in Poincar\'e coordinates $z = e^{-\rho/R}$,  attached to the brane located at $z=1$.
}
\label{fig:stringprofile}
\end{figure}

\begin{figure*}[t]
\begin{center}
\includegraphics[width=10cm]{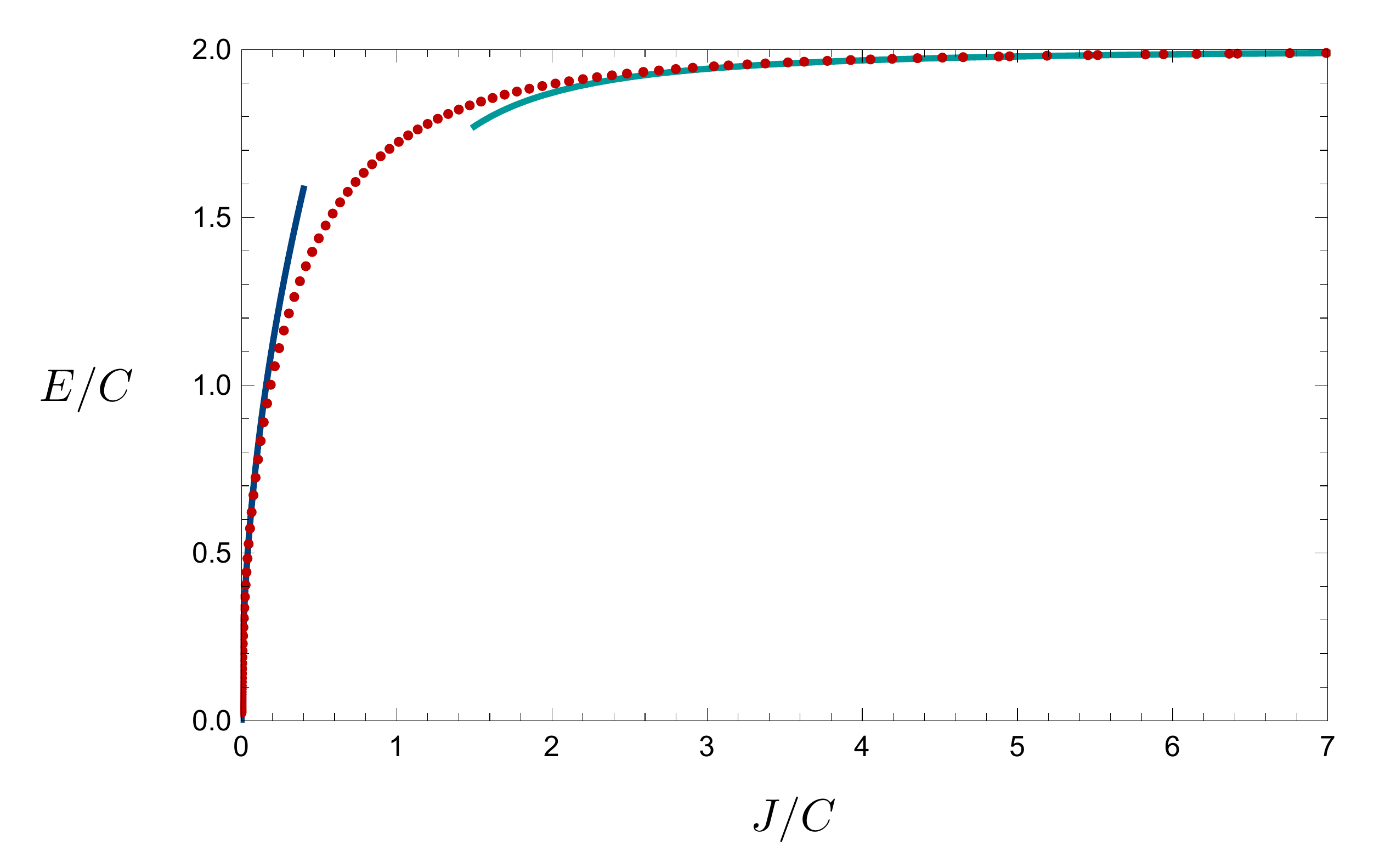}\\
\includegraphics[width=8cm]{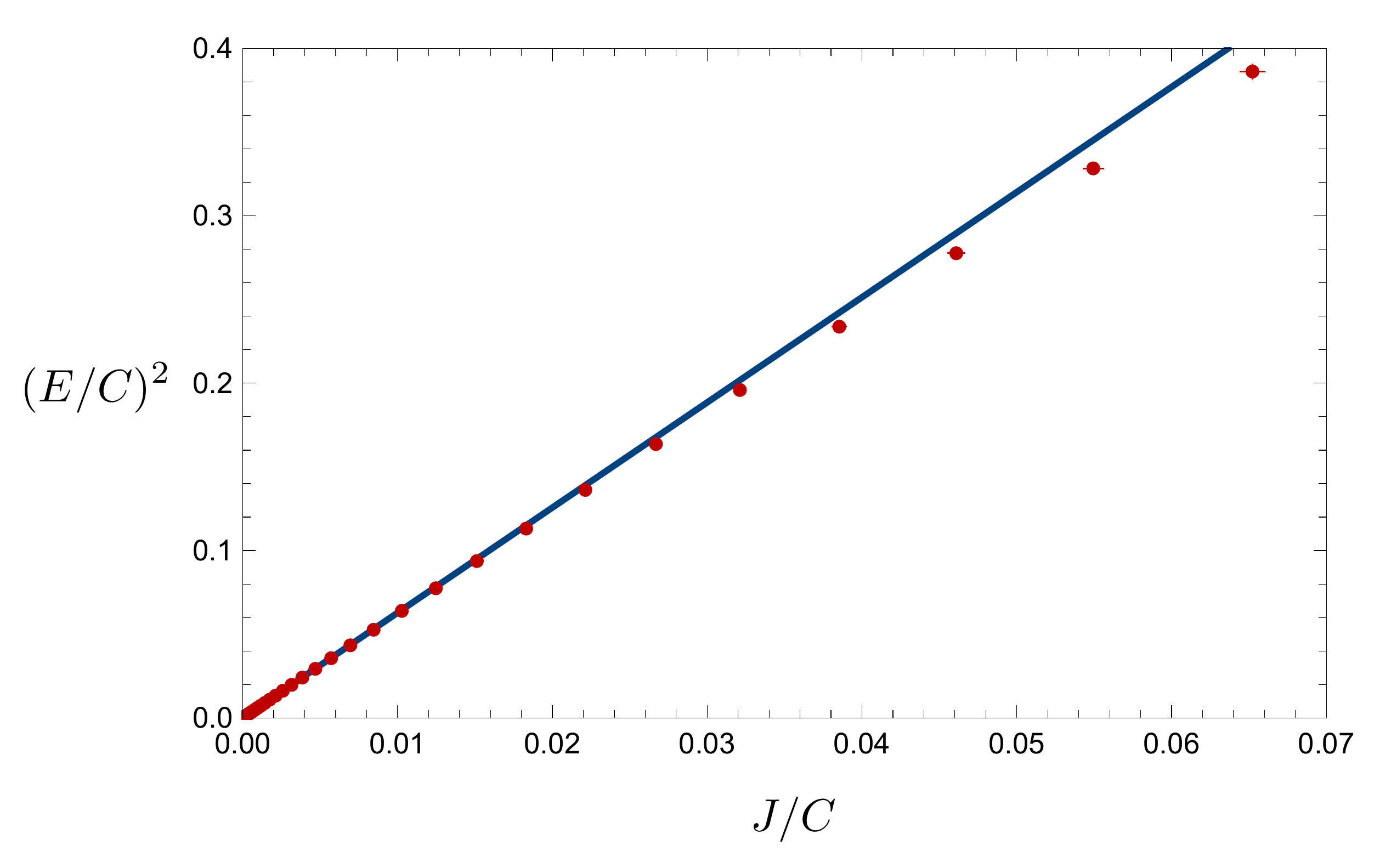}~~~~~~
\includegraphics[width=8cm]{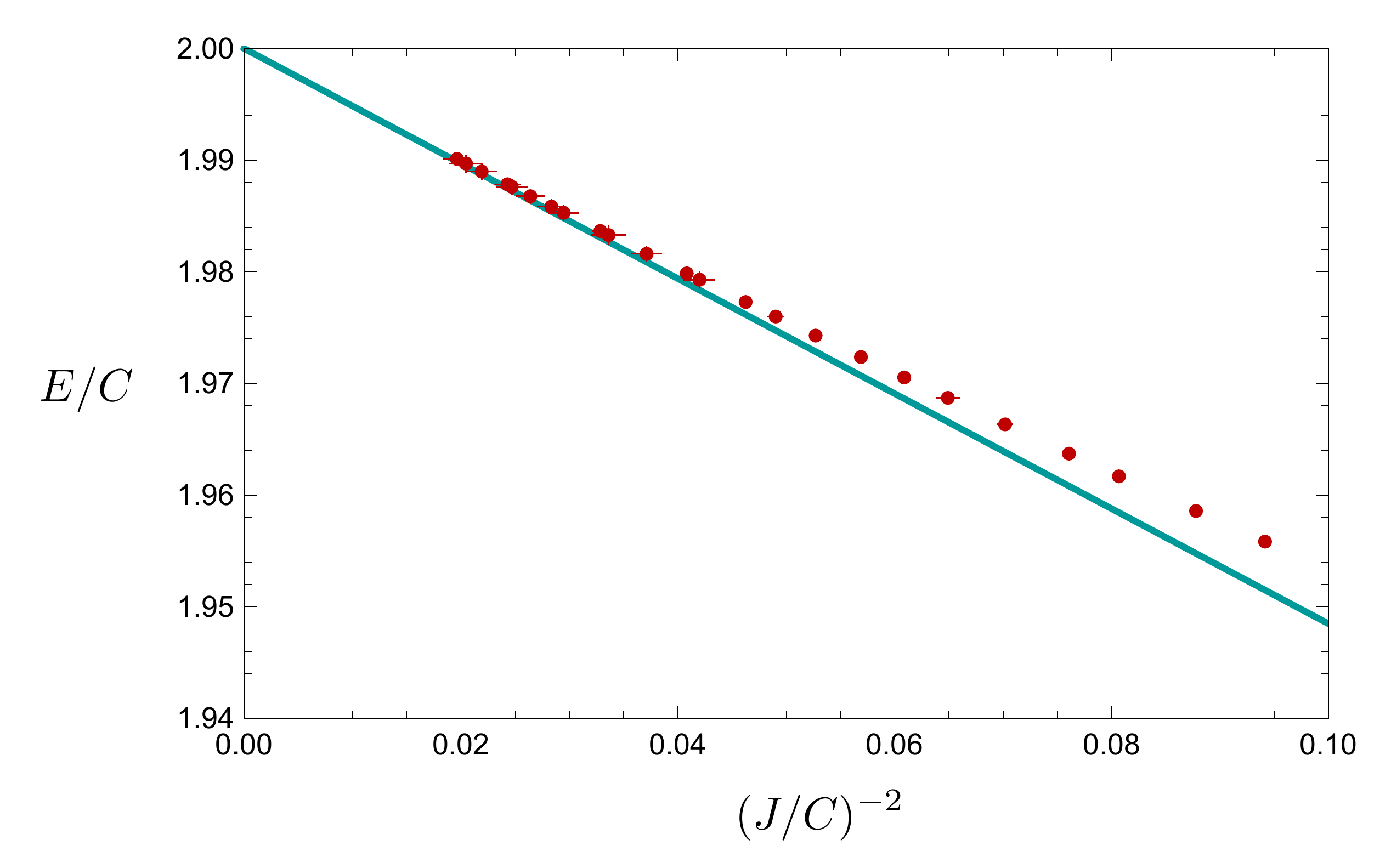}
\end{center}
\vspace{-0.5cm}
\caption{Numerical computation of the spectrum of the rotating string, for fixed values of $\omega$ (red points), for $\log \omega$ ranging from $-6$ to $4.9$ in increments of $0.1$ (top figure).
We write $E$ and $J$ in units of $C = R^2/2\pi \alpha'$. The asymptotic low- and high-energy analytic expressions of Eqs.~\nref{eq:lowJ} and \nref{eq:highJ} are plotted in blue and teal, respectively.
In the second row is a numerical check, with error bars plotted, of the approach to the asymptotic forms for (left) low  \nref{eq:lowJ} and (right) high~\nref{eq:highJ} energies.
}
\label{fig:spectrum}
\end{figure*}

Let us take as an ansatz a spinning string parameterized by target-space variables with dependences $(t(\tau),r(\sigma),\theta(\tau),z(\sigma))$, defining polar coordinates on ${\rm d}x_d^2$.
The equations of motion
require $t(\tau)$ and $\theta(\tau)$ to both be linear, so without loss of generality we write $t = \tau$ and $\theta = \omega\tau$.
Then the remaining nontrivial Virasoro constraint becomes
\be
 r'^2 + \omega^2 r^2 + z'^2 = 1. \label{eq:Vconstraint}
\ee
The $r$- and $z$-components of the $X^\mu$ equation of motion are equivalent on Eq.~\eqref{eq:Vconstraint}:
\be
 z z'' + 2 r'^2 = 0.\label{eq:EOM}
\ee

We numerically solve Eqs.~\eqref{eq:Vconstraint} and \eqref{eq:EOM} with the conditions $z(0)=z_0$, $z'(0)=0$, and $r(0)=0$, where $z_0$ is fixed by requiring $r'(\sigma_{\rm b}) = 0$, where $\sigma_{\rm b}$ is defined via $z(\sigma_{\rm b}) = 1$. We plot the solutions in Fig.~\ref{fig:stringprofile}.
Given a solution $(r(\sigma),z(\sigma))$, the energy and angular momentum are given by
\be 
E = 2C \int_0^{\sigma_{\rm b}} z^{-2}{\rm d}\sigma ~~~ {\rm and} ~~~ J = 2C \int_0^{\sigma_{\rm b}} \omega r^2 z^{-2} {\rm d}\sigma, ~~~~~~~C ={ R^2 \over 2 \pi \alpha'} ,
\ee
defining $E$ as conjugate to $t$. (To convert back to the energy in the main text, one simply divides by $R$.)
We plot the numerical results in  Fig.~\ref{fig:spectrum}.

At large $\omega$, $z$ approaches a constant at $z=1$, and the Virasoro constraint implies that $r\rightarrow \omega^{-1} \sin \omega\sigma$, so $\sigma_{\rm b} \rightarrow \pi/2\omega$.\footnote{Strictly speaking, we must include corrections around the $\omega\rightarrow\infty$ limit in order to solve the equation of motion~\eqref{eq:EOM}. For example, we can take $r = \frac{1}{\omega}\sin \omega\sigma$ as before, but $z = 1 - \frac{1}{8\omega^2}[4(\omega\sigma)^2 -\pi^2 - 4\cos^2 \omega\sigma]$. In scaleless units $\bar\sigma = \omega\sigma$, one then finds that both Eqs.~\eqref{eq:Vconstraint} and \eqref{eq:EOM} are satisfied to $O(\omega^{-2})$.
}
Thus, in the large-$\omega$ limit, we have $E/C =\pi/\omega$ and $J/C = \pi/2\omega^2$, so we obtain the flat-space formula for a rotating string given in Eq.~\eqref{eq:lowJ}.

\medskip
 
\bibliographystyle{utphys-modified}
\bibliography{AccumulationPoint.bib}

\end{document}